\documentclass[prd,aps,twocolumn,floatfix,superscriptaddress,preprintnumbers]{revtex4-2}

\usepackage{amsmath,mathtools,cases,slashed,bm}
\usepackage{color}
\usepackage[colorlinks=true,citecolor=blue,linkcolor=blue,urlcolor=blue]{hyperref}
\usepackage[dvipsnames]{xcolor}
\usepackage{amssymb}
\usepackage{tabularx}
\usepackage{physics}
\usepackage{hhline}
\usepackage[normalem]{ulem} 

\newcommand{\comment}[1]{}
\newcommand{\fref}[1]{Fig.~\ref{#1}}

\newcommand{\eref}[1]{Eq.~(\ref{#1})}

\newcommand{\tref}[1]{Table~\ref{#1}}
\newcommand{\diff}[1]{\mathrm{d}#1}
\newcommand{\chired}{{\chi^2_{\rm red}}}
\newcommand{\chirede}{{\chi^2_{{\rm red},e}}}
\newcommand{\ubar}{{\bar{u}}}
\newcommand{\dbar}{{\bar{d}}}
\newcommand{\sigDY}{\sigma^{\mbox{\tiny DY}}}
\newcommand{\sighatDY}{\hat\sigma^{\mbox{\tiny DY}}}
\newcommand{\sigW}{\sigma^{\mbox{\tiny $W$}}}
\newcommand{\sigWp}{\sigma^{\mbox{\tiny $W^+$}}}
\newcommand{\sigWm}{\sigma^{\mbox{\tiny $W^-$}}}
\newcommand{\sighatW}{\hat\sigma^{\mbox{\tiny $W$}}}

\begin{document}
\preprint{JLAB-THY-21-3491}

\title{Bayesian Monte Carlo extraction of sea asymmetry with SeaQuest and STAR data}

\author{C. Cocuzza}
\affiliation{Department of Physics, SERC, Temple University, Philadelphia, Pennsylvania 19122, USA}
\author{W. Melnitchouk}
\affiliation{Jefferson Lab,
	     Newport News, Virginia 23606, USA \\
        \vspace*{0.2cm}
        {\bf Jefferson Lab Angular Momentum (JAM) Collaboration
        \vspace*{0.2cm} }}
\author{A. Metz}
\affiliation{Department of Physics, SERC, Temple University, Philadelphia, Pennsylvania 19122, USA}
\author{N. Sato}
\affiliation{Jefferson Lab,
	     Newport News, Virginia 23606, USA \\
        \vspace*{0.2cm}
        {\bf Jefferson Lab Angular Momentum (JAM) Collaboration
        \vspace*{0.2cm} }}
\date{\today}

\begin{abstract}
We perform a global QCD analysis of unpolarized parton distributions within a Bayesian Monte Carlo framework, including the new $W$-lepton production data from the STAR Collaboration at RHIC and Drell-Yan di-muon data from the SeaQuest experiment at Fermilab. 
We assess the impact of these two new measurements on the light antiquark sea in the proton, and the $\dbar-\ubar$ asymmetry in particular.
The SeaQuest data are found to significantly reduce the uncertainty on the $\dbar/\ubar$ ratio at large parton momentum fractions $x$, strongly favoring an enhanced $\bar d$ sea up to $x \approx 0.4$, in general agreement with nonperturbative calculations based on chiral symmetry breaking in QCD.
\end{abstract}

\maketitle

\section{Introduction}

Over the past five decades, there has been great interest in the structure of the nucleon sea, which encompasses quarks, antiquarks, and gluons that exist beyond the three valence quarks originally proposed as the building blocks of the nucleon~\cite{Gell-Mann:1964ewy, Zweig:1964ruk, Feynman:1973xc}.
In particular, high-energy scattering experiments and global QCD analyses of the data have now conclusively demonstrated a large difference in the momentum distributions of $\bar u$ and $\bar d$ antiquarks in the proton~\cite{Geesaman:2018ixo}.
The result cannot be explained perturbatively through the splitting of gluons into quark-antiquark pairs~\cite{Ross:1978xk}, and requires nonperturbative mechanisms, such as dynamical chiral symmetry breaking and the pion cloud of the nucleon~\cite{Thomas:1983fh, Schreiber:1991qx, Henley:1990kw, Kumano:1991em, Melnitchouk:1991ui, Speth:1996pz, Kumano:1997cy, Melnitchouk:1998rv, Pobylitsa:1998tk, Chang:2014jba, Alberg:2017ijg}, or dynamics related to the Fermi-Dirac statistics of quarks and the Pauli exclusion principle~\cite{Field:1976ve, Schreiber:1991tc, Bourrely:1994sc, Bourrely:1994nm, Steffens:1996bc, Broadhurst:2004jx}.
More recently, exploratory studies have been made in extracting information on the isovector sea quark distributions directly from lattice QCD calculations~\cite{Ji:2013dva, Ma:2014jla, Orginos:2017kos, Chen:2017mzz, Alexandrou:2021oih}.

The first experimental indications of a light-quark sea asymmetry came from the CFS group at Fermilab in 1981~\cite{Ito:1980ev}. 
Measuring the Drell-Yan process, where a quark and antiquark from colliding hadrons annihilate into a virtual photon that subsequently decays into a lepton-antilepton pair, they found that the $\dbar$ distribution, integrated over parton momentum fraction $x$, was larger than the integrated $\ubar$ distribution. 
The first high-precision experimental evidence for a sea asymmetry came a decade later from the New Muon Collaboration (NMC) at CERN~\cite{NewMuon:1991hlj, NewMuon:1993oys}, which used measurements of inclusive deep-inelastic scattering (DIS) on hydrogen and deuterium to test the Gottfried sum rule~\cite{Gottfried:1967kk} and determine that the integral
    $\int_0^1 \diff x \big[ \dbar(x)-\ubar(x) \big]$ 
must be positive.
Further evidence was provided by the NA51 Collaboration at CERN~\cite{NA51:1994xrz} using the Drell-Yan process, and indications for a nonzero asymmetry were also found by the HERMES Collaboration~\cite{HERMES:1998uvc} in semi-inclusive DIS of charged pions.

The most conclusive evidence for an excess of $\dbar$ over $\ubar$ was provided in 1998 by the Fermilab E866 (NuSea) experiment~\cite{Webb:2003bj, NuSea:1998kqi, NuSea:2001idv}, which measured Drell-Yan lepton-pair production cross sections in proton-proton ($pp$) and proton-deuteron ($pD$) scattering.
At kinematics where the parton momentum fraction of the beam, $x_1$, is much greater than that of the target, $x_2$, one finds that the ratio of $pD$ to $pp$ cross section, at leading order in the strong coupling, provides direct sensitivity to the $\dbar/\ubar$ ratio~\cite{Ellis:1990ti},
\begin{align}
\frac{\sigDY_{pD}}{2\sigDY_{pp}} \bigg|_{x_1 \gg x_2}
\approx \frac12 \bigg[ 1 + \frac{\bar{d}(x_2)}{\bar{u}(x_2)} \bigg].
\label{e.DYrat}
\end{align}

While the NuSea experiment probed the asymmetry up to momentum fractions $x \approx 0.3$, the subsequent Fermilab E906 (SeaQuest) experiment~\cite{SeaQuest:2021zxb} extended the range up to $x \approx 0.4$, finding some tensions with the NuSea result in the high-$x$ region.
The NuSea data had suggested a significant fall in the $\dbar/\ubar$ ratio for $x \gtrsim 0.3$, albeit with large uncertainties, which is difficult to accommodate in many of the nonperturbative models~\cite{Thomas:1983fh, Schreiber:1991qx, Henley:1990kw, Kumano:1991em, Melnitchouk:1991ui, Speth:1996pz, Kumano:1997cy, Melnitchouk:1998rv, Pobylitsa:1998tk, Chang:2014jba, Alberg:2017ijg, Field:1976ve, Schreiber:1991tc, Bourrely:1994sc, Bourrely:1994nm, Steffens:1996bc}.
The SeaQuest experiment was partially motivated to verify this behavior at large $x$.

An alternative method for extracting the $\dbar-\ubar$ asymmetry involves $W$-lepton production in hadronic collisions, whereby a quark and antiquark annihilate into a $W$ boson that decays into a detected lepton and a neutrino.  This has been measured in $p\bar p$ scattering at the Tevatron
and in $pp$ collisions at the Large Hadon Collider (LHC)
and the Relativistic Heavy Ion Collider (RHIC). 
Taking the ratio of the $W^+$ and $W^-$ cross sections for $pp$ reactions, at leading order in perturbative QCD one has
\begin{align} 
\frac{\sigWp_{pp}}{\sigWm_{pp}}
\approx 
\frac{u(x_1) \bar{d}(x_2) + u(x_2) \bar{d}(x_1)}
{d(x_1) \bar{u}(x_2) + d(x_2) \bar{u}(x_1)}.
\label{e.Wrat}
\end{align}
The most recent high-precision data from the STAR Collaboration at RHIC~\cite{STAR:2020vuq} on the ratios of $W$-lepton production cross sections have not yet been included in global QCD analyses.
Since RHIC has a lower center-of-mass energy of $\sqrt{S} = 0.51$~TeV compared to the Tevatron ($\sqrt{S} = 1.96$~TeV) and the LHC ($\sqrt{S} \geq 7$~TeV), these data are sensitive to parton distributions at higher values of~$x$, which can potentially provide information on the distributions of light sea quarks in this difficult to measure region.

In this paper, we present the results of a global QCD analysis using the JAM Bayesian Monte Carlo framework, including the new measurements from SeaQuest~\cite{SeaQuest:2021zxb} and STAR~\cite{STAR:2020vuq}.
By combining these data with other hadron collider data from Fermilab and the LHC, and with DIS data from fixed-target and collider experiments, we are able to place strong constraints on the light-quark PDFs and assess the impact of the new data on the antiquark asymmetry $\dbar-\ubar$.

In Sec.~\ref{sec.framework} we begin by reviewing the theoretical framework used in this analysis, including a summary of the most relevant factorization formulas.
In Sec.~\ref{sec.bayes} we review the methodology employed for the analysis,  as well as the choice of parametrization for the parton distribution functions (PDFs).
The quality of the fit is described in Sec.~\ref{sec.fit}, where we compare the fitted cross sections with the new Drell-Yan and $W$-lepton data.
The results for the PDFs are presented in Sec.~\ref{sec.pdfs}, where we in particular assess the impact of the various datasets on the shape of the $\dbar-\ubar$ asymmetry and its uncertainties.
We also compare our extracted asymmetry with nonperturbative calculations based on chiral symmetry breaking in QCD.
Finally, in Sec.~\ref{sec.outlook} we summarize our findings and discuss their implications for our understanding of the sea content of the nucleon.

\section{Theoretical framework}
\label{sec.framework}

Our theoretical framework is based on fixed order collinear factorization for various high-energy scattering processes involving spin-averaged PDFs, such as inclusive DIS, Drell-Yan lepton-pair production, and weak boson and jet production.
Since the new datasets that provide sensitivity to the antiquark distributions in this analysis involve Drell-Yan and weak boson production, we will focus on these observables in this section.

The double differential Drell-Yan lepton-pair production cross section can be written in terms of convolutions of the PDFs in the colliding hadrons with short-distance partonic cross sections $\sighatDY_{ab}$~\cite{Becher:2007ty},
\begin{align}
&\frac{\diff^2\sigDY}{ \diff M_{\ell\ell}^2\, \diff Y} 
= \frac{4\pi\alpha^2}{9 S M_{\ell\ell}^2}\sum_{ab}
  \int \diff x_1 \int \diff x_2  \\
& \qquad \times 
  f_a(x_1,\mu_F)\, f_b(x_2,\mu_F)\, \sighatDY_{ab}(x_1,x_2,S,M_{\ell\ell},\mu_R,\mu_F),
  \notag
\end{align}
where $\alpha$ is the electromagnetic coupling, $S$ is the invariant mass squared of the reaction, and $\mu_R$ and $\mu_F$ are the renormalization and factorization scales, respectively.
We write the cross section as differential in the invariant mass of the lepton pair, $M_{\ell\ell}^2$, and the rapidity of the lepton pair in the center-of-mass frame, $Y$.
The sum over the quark flavors $a, b$ runs over all partonic channels that can contribute to the Drell-Yan process, for which the scale is set to
    $\mu_R = \mu_F = M_{\ell\ell}$.
The partonic cross sections $\sighatDY_{ab}$ are computed at next-to-leading order (NLO) in the strong coupling $\alpha_s(\mu_R)$, with the NLO expressions taken from Ref.~\cite{Becher:2007ty}.
For $pp$ scattering, the PDFs $f_{a,b}$ are those in the proton.
For proton-deuteron scattering, the $x$ of the parton in the deuteron target is small enough that it is reasonable to approximate the $pD$ cross section by a simple sum of proton and neutron cross sections~\cite{Ehlers:2014jpa},
    $\sigma_{pD} \approx \sigma_{pp} + \sigma_{pn}$,
with the PDFs in the neutron related to those in the proton through isospin symmetry.

For $W$-lepton production, the double differential cross section is given by~\cite{Ringer:2015oaa}
\begin{align}
\label{e.W-lep}
& \frac{\diff^2 \sigW}{ \diff p_T^\ell\, \diff \eta_\ell}
= \frac{2}{p_T^\ell}
  \sum_{ab}\int_{VW}^V \diff v \int_{VW/v}^1 \diff w   \\
& \qquad \times
  x_1 f_a(x_1,\mu_F)\, x_2 f_b(x_2,\mu_F)\,
 \sighatW_{ab}(v,w,S,\mu_R,\mu_F),                      \notag
\end{align}
which is differential in the outgoing lepton's pseudorapidity, $\eta_\ell$, and its transverse momentum, $p_T^\ell$.
Labeling the momenta of the incoming hadrons by $P_1$ and $P_2$, we define the Mandelstam invariants $T \equiv (P_1-p_\ell)^2$ and $U \equiv (P_2-p_\ell)^2$, which then allows us to introduce the variables $V \equiv 1 + T/S$ and $W \equiv -U/(S+T)$. 
Defining $v$ and $w$ as the partonic analogues of $V$ and $W$, we can write for the parton momentum fractions $x_1 = VW/vw$ and $x_2 = (1-V)/(1-v)$.
The sum over the quark flavors $a,b$ runs over all partonic channels that can contribute to $W$-lepton production, for which the renormalization and factorization scales are chosen to be the mass of the $W$ boson, $\mu_R = \mu_F = M_W$.  
The partonic cross sections $\sighatW_{ab}$ are again computed at NLO in the strong coupling $\alpha_s(\mu_R)$, with the NLO expressions used here taken from Ref.~\cite{Ringer:2015oaa}.

In practice, the $W$-lepton production cross sections are usually integrated over a specified range of $p_T^\ell$ (for example, $p_T^\ell > 25$~GeV for the STAR data), and the cross section is given as differential in the pseudorapidity $\eta_\ell$.
Note that the theoretical formalism that leads to Eq.~(\ref{e.W-lep}) above applies to final states that are fully inclusive in the transverse momentum of the undetected neutrino,~$p_T^\nu$.
This formalism therefore cannot be used directly to calculate the $W$-lepton asymmetries from the D0, CDF, and ATLAS collaborations, which place cuts on $p_T^\nu$ and would require extending the current theoretical framework to implement such phase space cuts~\cite{Ebert:2020dfc}.

To avoid this ambiguity, we use the fully inclusive reconstructed $W$-boson data from D0 and CDF instead of the $W$-lepton data.
Our analysis does not include data from ATLAS, but includes the fully inclusive data from the CMS collaboration, which cover the same kinematic region with similar uncertainties.
We can thus retain most of the information provided by these observables, despite the exclusion of the non-fully inclusive $W$-lepton data from CDF, D0, and ATLAS.

For the DIS theory, we include corrections that are known to be necessary to accurately describe the high-$x$ region. 
These include target mass corrections, as described by Moffat {\it et al.} \cite{Moffat:2019qll}, parameterized higher twist corrections, and nuclear corrections for deuterium data. 
Further details regarding these corrections will be discussed in a future publication~\cite{Cocuzza:2021DIS}.

The scale dependence of the PDFs is determined according to the DGLAP evolution equations~\cite{Gribov:1972ri, Dokshitzer:1977sg, Altarelli:1977zs}, with the PDFs and $\alpha_s$ evolved according to their renormalization group equation (RGE) at next-to-leading logarithmic accuracy with the boundary condition $\alpha_s(M_Z)=0.118$. 
For light and for heavy quarks, the PDFs are evolved using the zero-mass variable-flavor-number scheme.
The values of the heavy quark mass thresholds for the evolution of the PDFs and $\alpha_S$ are taken from the PDG values $m_c=1.28$~GeV and $m_b=4.18$~GeV in the $\overline{\textrm{MS}}$ scheme~\cite{ParticleDataGroup:2018ovx}. 

\section{Bayesian Inference}
\label{sec.bayes}

Our PDF extraction procedure is based on Bayesian inferences using Monte Carlo techniques developed in previous JAM analyses~\cite{Sato:2016tuz, Sato:2016wqj, Ethier:2017zbq, Sato:2019yez, Moffat:2021dji}. 
We parametrize the PDFs at the input scale $\mu_0^2 = m_c^2$ using a generic template function of the form
\begin{align}
f(x,\mu_0^2) 
= \frac{N}{\cal M}\, x^{\alpha}(1-x)^{\beta}(1+\gamma \sqrt{x} + \eta x),
\label{e.template}
\end{align}
where $\bm{a} = \{ N, \alpha, \beta, \gamma, \eta \}$ is the set of parameters to be inferred, and 
    ${\cal M} = {\rm B}[\alpha+2,\beta+1]
              + \gamma {\rm B}[\alpha+\frac52,\beta+1]
              + \eta {\rm B}[\alpha+3,\beta+1]$ 
normalizes the function to the second moment to maximally decorrelate the normalization and shape parameters.
To characterize the nucleon valence region and discriminate it from the sea components, we parametrize the light-quark and strange PDFs according to
\begin{align}
u      &= u_v + \bar{u},    \hspace{0.75cm}
d       = d_v + \bar{d},    \notag\\         
\bar{u}&= S_1 + \bar{u}_0,  \hspace{0.60cm}
\bar{d} = S_1 + \bar{d}_0,  
\label{e.param}             \\
s      &= S_2 + s_0,        \hspace{0.65cm}
\bar{s} = S_2 + \bar{s}_0,  \notag
\end{align}
where the dependence on $x$ and the scale $\mu_0^2$ has been suppressed for convenience.
The input quark distributions $u_v$, $d_v$, $\bar{u}_0$, $\bar{d}_0$, $s_0$, and $\bar{s}_0$, as well as the gluon distribution $g$, are parametrized individually as in \eref{e.template}. 
For the sea quark PDFs, the additional functions $S_1$ and $S_2$ are also parametrized via \eref{e.template}, and are designed to allow a more singular small-$x$ behavior compared to the valence distributions by restricting the corresponding $\alpha$ parameter to more negative values.
Letting $S_1 \neq S_2$ allows for different small-$x$ behaviors for the light sea quarks and the strange quarks.

The normalization parameter $N$ for the gluon distribution is fixed by the momentum sum rule, while the corresponding normalization parameters for $u_v$, $d_v$, and $s_0$ are fixed by the valence number sum rules.
For the $u_v$, $d_v$, $g$, $\ubar_0$, and $\dbar_0$ distributions, the parameters $\gamma$ and $\eta$ in \eref{e.template} are included to allow sufficient flexibility.
We have verified that including these parameters for $S_1$, $S_2$, $s_0$, or $\bar{s}_0$ does not lead to significant changes to the final results, so for these distributions the $\gamma$ and $\eta$ parameters are set to zero.

Our Bayesian analysis consists of sampling the posterior distribution given by
\begin{align}
\mathcal{P}(\bm{a}|{\rm data})
= \mathcal{L}(\bm{a},{\rm data})\, \pi(\bm{a}),
\end{align}
with a likelihood function of Gaussian form,
\begin{align}
\mathcal{L}(\bm{a},{\rm data}) 
= \exp\Big( -\frac12 \chi^2(\bm{a},{\rm data}) \Big),
\label{e.likelihood}
\end{align}
and a flat prior function $\pi(\bm{a})$ that vanishes in regions where the parameters $\bm{a}$ give unphysical PDFs.
The $\chi^2$ function in (\ref{e.likelihood}) is defined as
\begin{align}
\chi^2(\bm{a}) &= \sum_{i,e} 
\bigg( 
\frac{d_{i,e} - \sum_k r_e^k \beta_{i,e}^k - T_{i,e}(\bm{a})/N_e}{\alpha_{i,e}} \bigg)^2
\notag\\
&\quad
+ \sum_k \big(r_e^k\big)^2 + \sum_e \bigg( \frac{1-N_e}{\delta N_e} \bigg)^2,
\end{align}
where $d_{i,e}$ is the experimental data point $i$ from dataset $e$, and $T_{i,e}$ is the corresponding theoretical value.
All uncorrelated uncertainties are added in quadrature and labeled by $\alpha_{i,e}$, while $\beta_{i,e}^k$ represents the $k$-th source of point-to-point correlated systematic uncertainties for the $i$-th data point weighted by $r_e^k$.
The latter are optimized per values of the parameters $\bm{a}$ via $\partial \chi^2/\partial r_{e}^k=0$.
We include normalization parameters $N_e$ for each dataset $e$ as part of the posterior distribution per data set, with a Gaussian penalty controlled by the experimentally quoted normalization uncertainties $\delta N_e$.

The posterior distribution is sampled via data re-sampling, whereby multiple maximum likelihood optimizations are carried out by adding Gaussian noise with width $\alpha_{i,e}$ to each data point across all data sets.
The resulting ensemble of parameter samples $\{\bm{a}_k; k=1, \ldots, n\}$ is then used to obtain statistical estimators for a given observable (function of PDFs) ${\cal O}({\bm{a}})$, such as the mean and variance,
\begin{subequations}
\label{e.EandVdef}
\begin{align}
{\rm E}[{\cal O}]
&= \frac{1}{n} \sum_k {\cal O}(\bm{a}_k),
\\
{\rm V}[{\cal O}]
&= \frac{1}{n} \sum_k \big[ {\cal O}(\bm{a}_k)-{\rm E}[{\cal O}] \big]^2.
\end{align}
\end{subequations}
The agreement between data and theory is assessed by defining the ``reduced'' $\chi^2$ for each dataset $e$ as
\begin{align}
\label{e.chired}
\chirede \equiv \frac{1}{N_{\rm dat}^e} 
\sum_i
\bigg( 
    \frac{d_{i,e}-{\rm E}\big[ \sum_k r_e^k \beta_{i,e}^k + T_{i,e}/N_e \big]}
    {\alpha_{i,e}}
\bigg)^2,
\end{align}
with $N_{\rm dat}^e$ the total number of data points for each experiment, and ${\rm E}[...]$ represents the mean theory as defined in \eref{e.EandVdef}.

\begin{figure*}[t]
\includegraphics[width=0.9\textwidth]{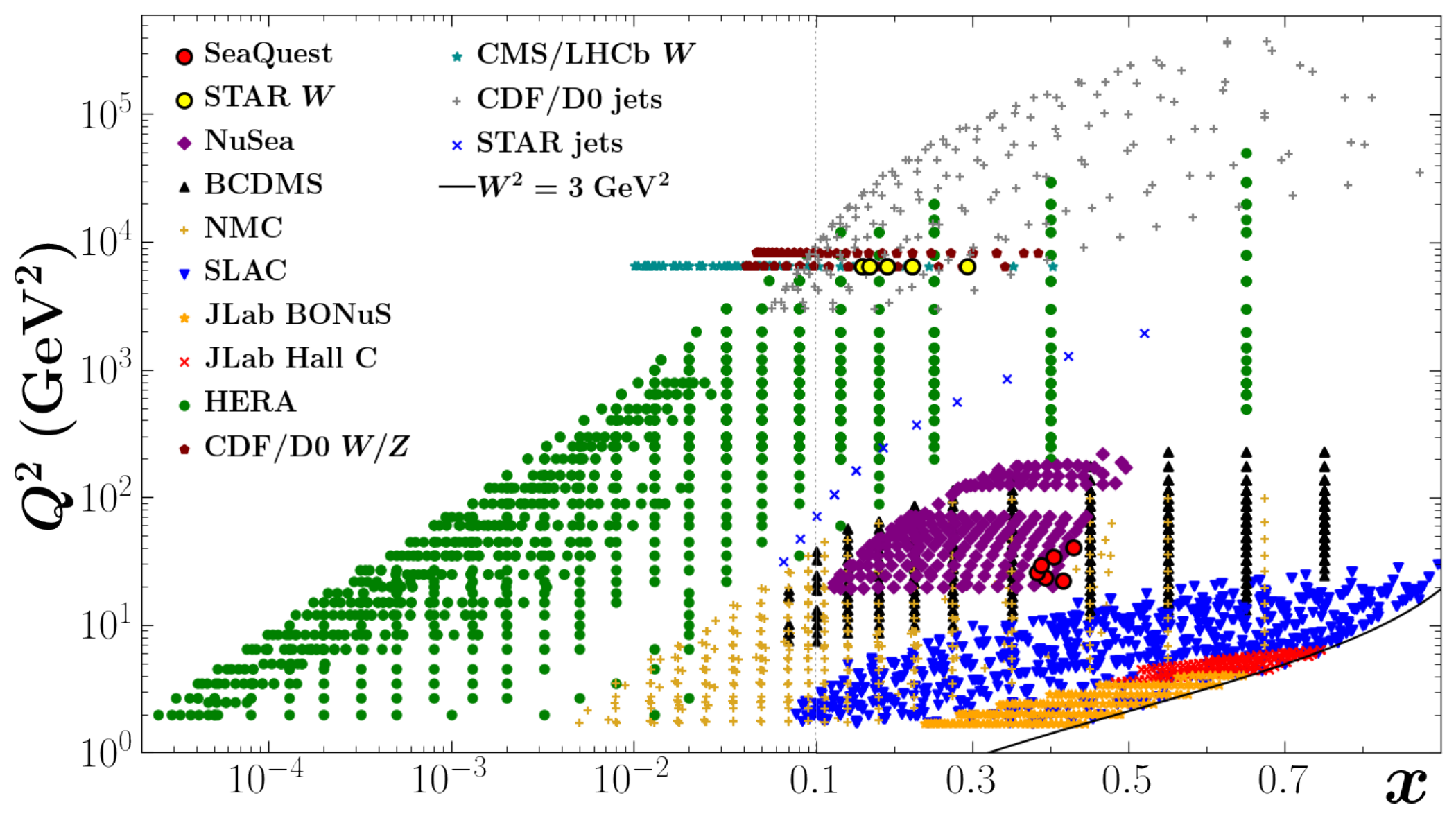}
\caption{Kinematic coverage of the datasets included in this analysis.  The variable $x$ represents Bjorken-$x$ for DIS and Feynman-$x$ for vector boson and jet production, while the scale $Q^2$ represents the four-momentum transfer squared for DIS, the mass squared of the intermediate boson for vector boson production, and the transverse momentum squared for jet production.  Also indicated is the DIS cut of $W^2 = 3$~GeV$^2$ (solid back line).}
\label{f.kin}
\end{figure*}

\begin{figure}[t]
\includegraphics[width=0.48\textwidth]{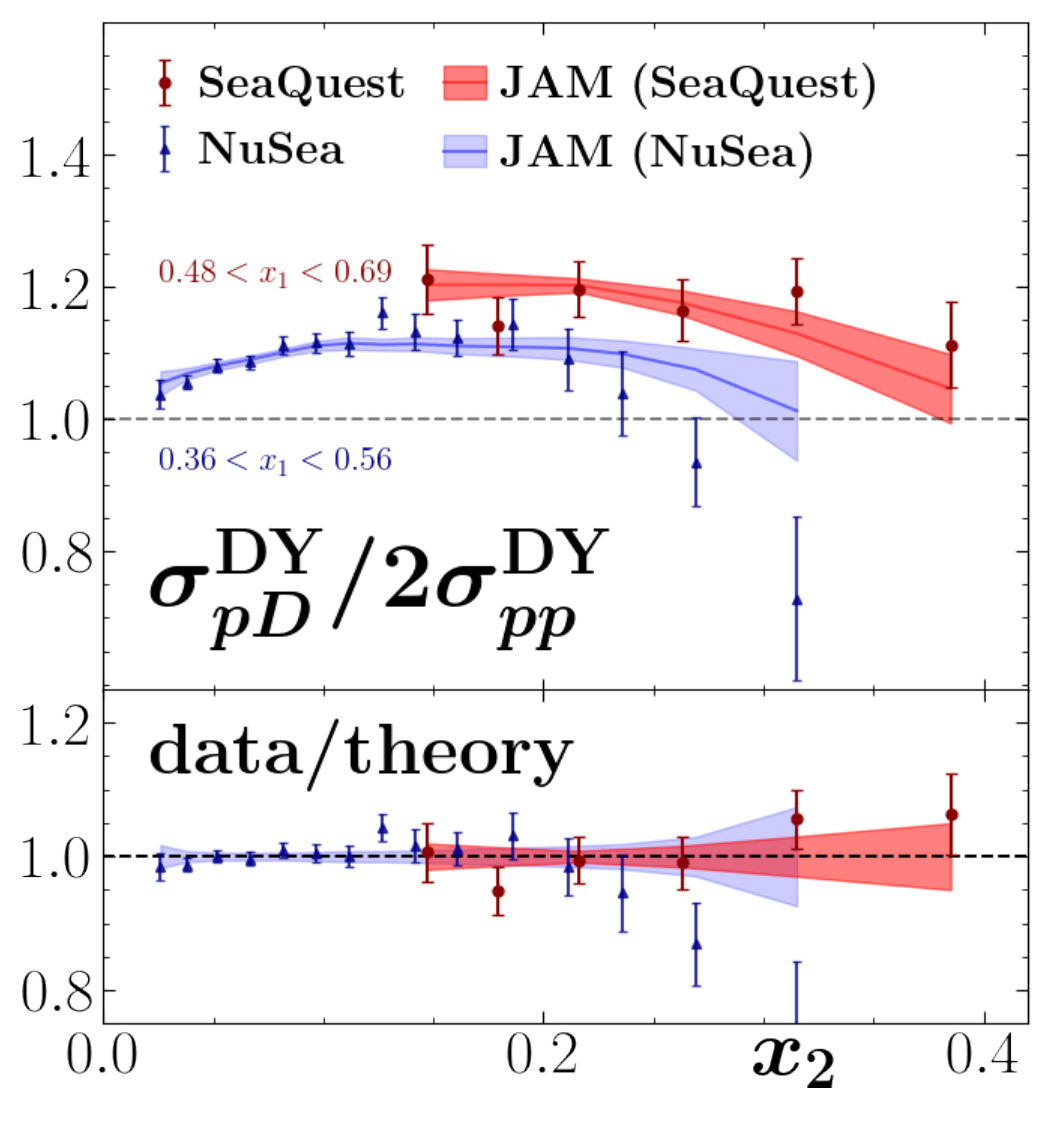}
\caption{Drell-Yan cross section ratio $\sigDY_{pD}/2\sigDY_{pp}$ from SeaQuest~\cite{SeaQuest:2021zxb} (red circles) and NuSea~\cite{NuSea:2001idv} (blue triangles) compared with the JAM results at their respective kinematics (red and blue 1$\sigma$ uncertainty bands), as a function of the target momentum fraction $x_2$, with the corresponding $x_1$ ranges indicated.  The ratio of data to the average theory is illustrated in the lower panel with 1$\sigma$ theoretical uncertainties at the SeaQuest and NuSea kinematics.}
\label{f.SQ}
\end{figure}
\begin{figure}[t]
\includegraphics[width=0.48\textwidth]{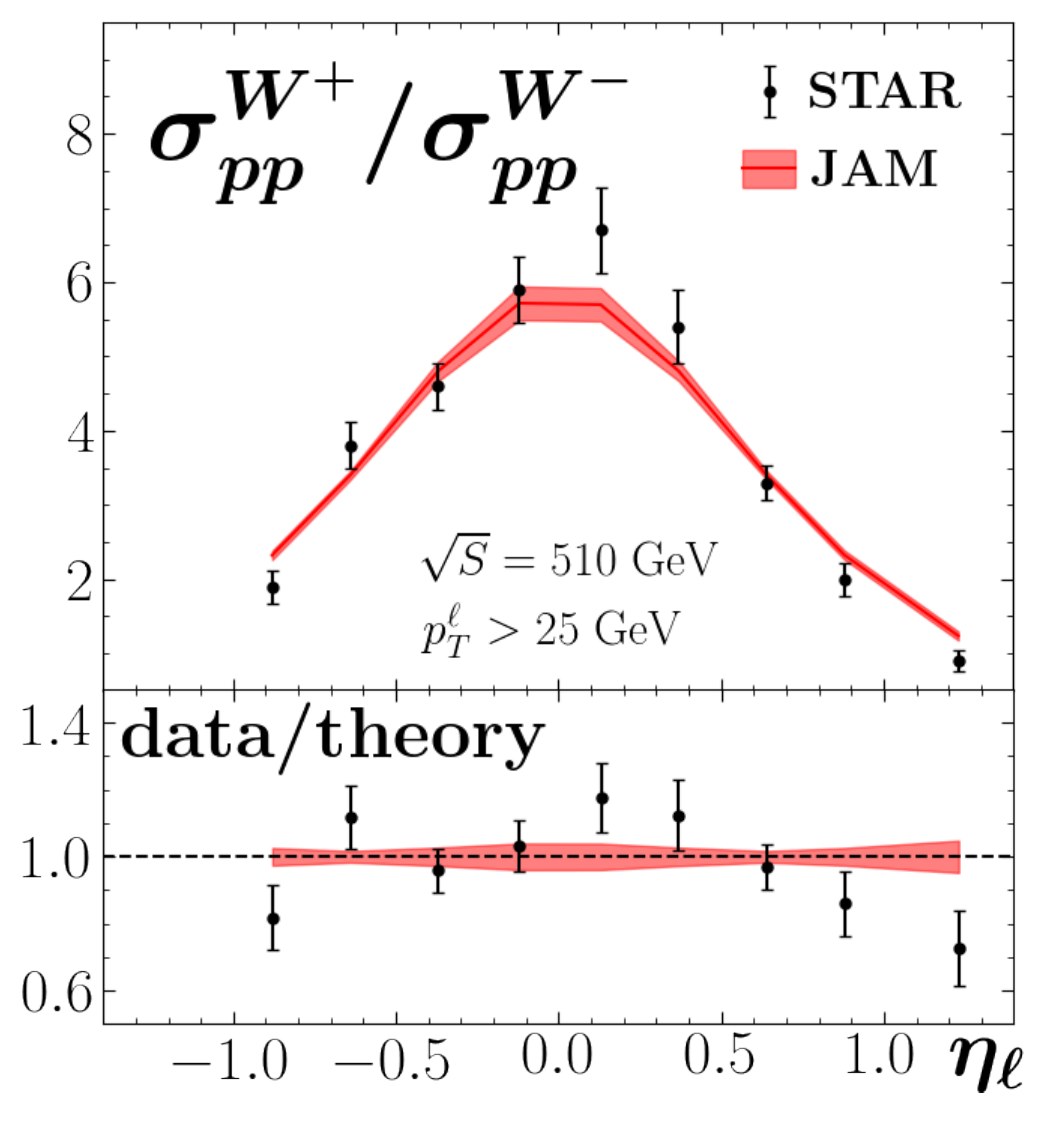}
\caption{$W$-lepton cross section ratio $\sigWp_{pp}/\sigWm_{pp}$ from STAR~\cite{STAR:2020vuq} (black circles) compared with the JAM fit (red 1$\sigma$ uncertainty band), as a function of the lepton pseudorapidity, $\eta_\ell$.  The ratio of data to average theory is shown in the lower panel together with the 1$\sigma$ theoretical uncertainty.}
\label{f.STAR}
\end{figure}

\begin{figure*}[t]
\includegraphics[width=0.95\textwidth]{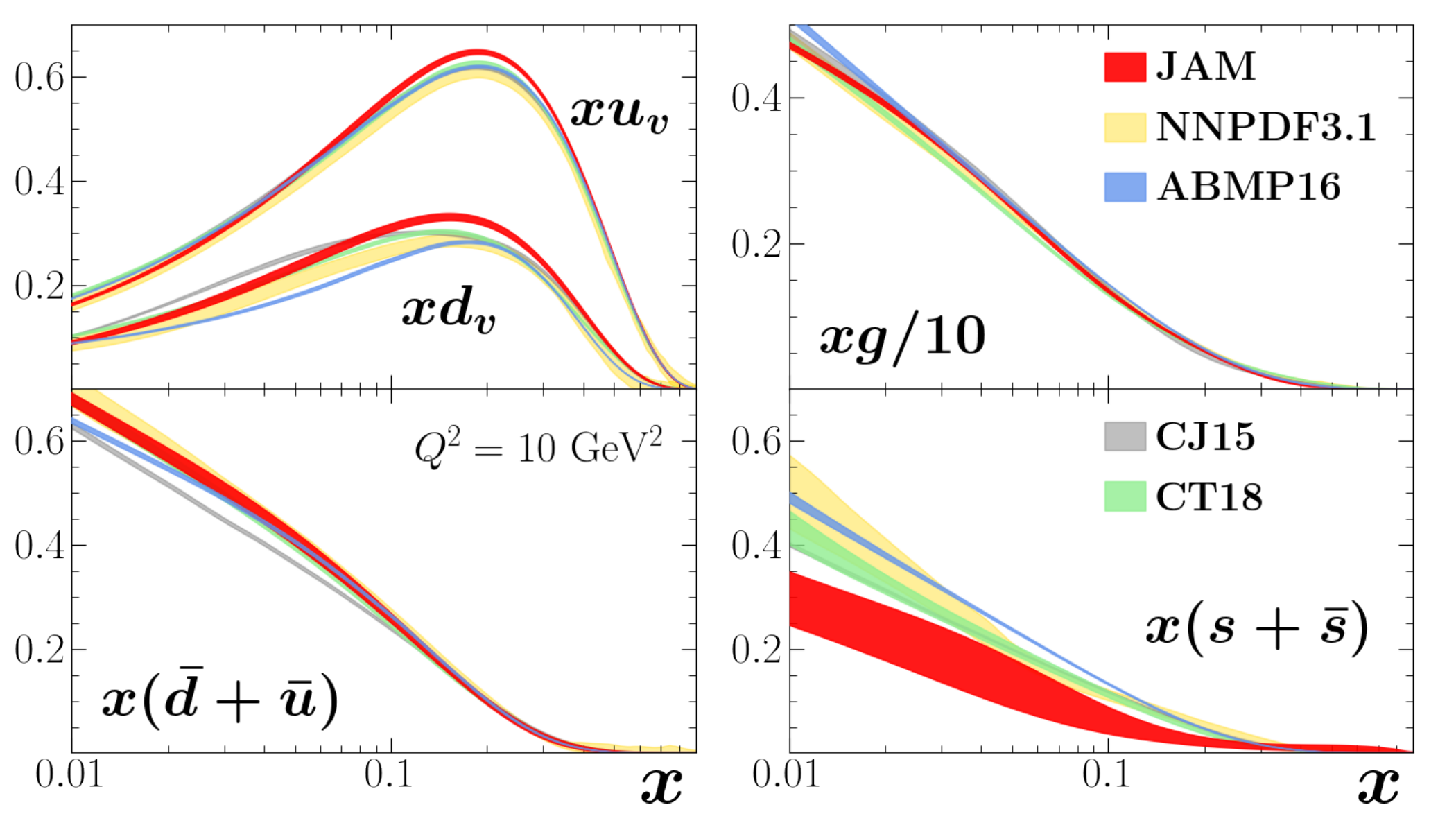}
\caption{Comparison of the JAM PDFs from this analysis (red bands) with NLO results from the NNPDF3.1~\cite{NNPDF:2017mvq} (gold), ABMP16~\cite{Alekhin:2017kpj} (blue), CJ15~\cite{Accardi:2016qay} (gray), and CT18~\cite{Hou:2019efy} (green) parametrizations at the scale $Q^2 = 10$~GeV$^2$, with the bands representing 1$\sigma$ uncertainty.  Note that $x$ multiplied by the PDFs are shown.}
\label{f.PDFs}
\end{figure*}
\begin{figure}[t]
\includegraphics[width=0.48\textwidth]{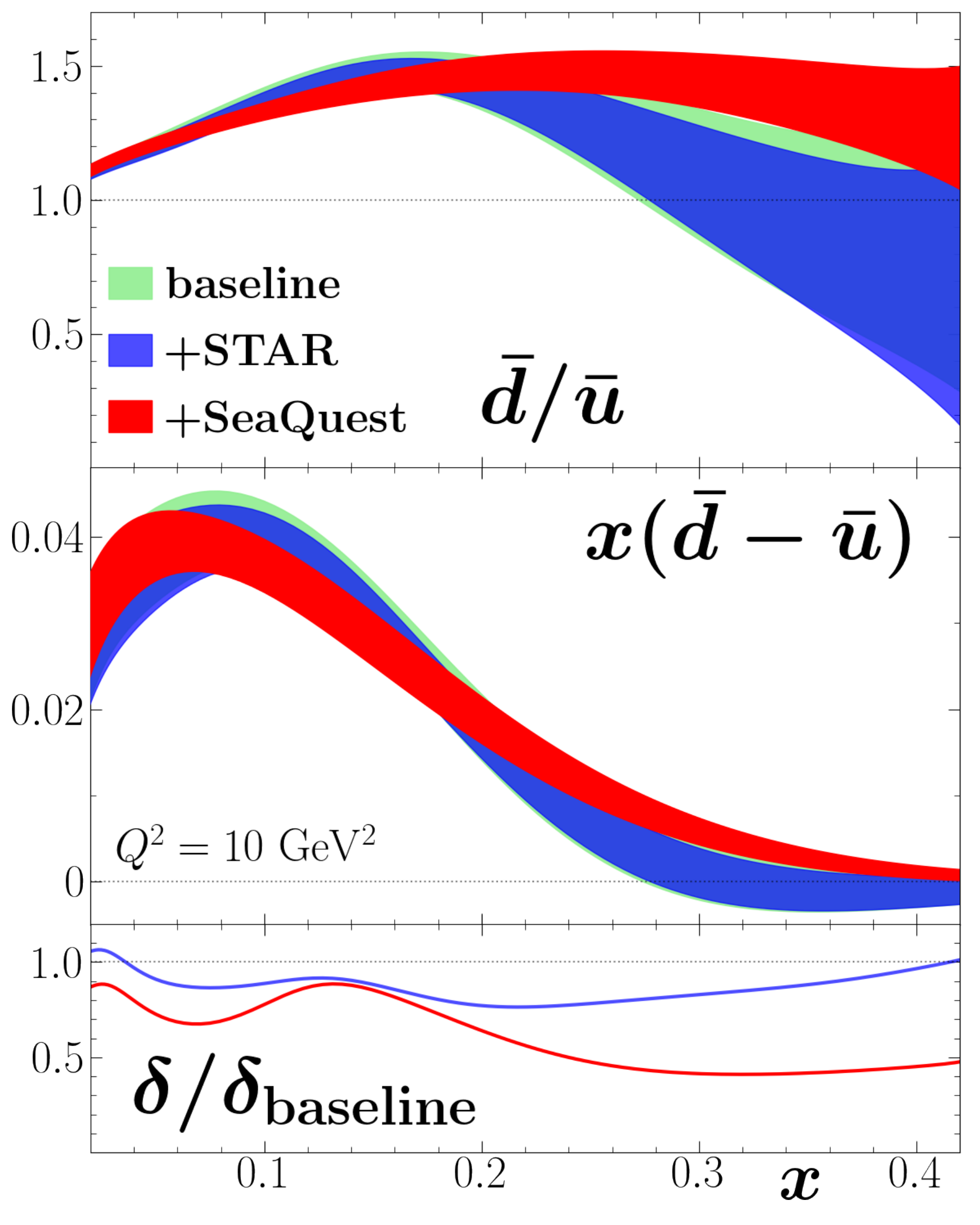}
\caption{Impact on the $\dbar/\ubar$ ratio (top panel) and the asymmetry $x(\dbar-\ubar)$ (middle panel) of the $W$-lepton data from STAR~\cite{STAR:2020vuq} (blue bands) and the Drell-Yan measurement from SeaQuest~\cite{SeaQuest:2021zxb} (red bands) relative to the ``baseline'' (green bands) which contains all data except these.  The STAR and SeaQuest data are added in succession.  The uncertainty on $\dbar/\ubar$ for these two scenarios normalized to that of the baseline are shown in the bottom panel.  All bands represent 1$\sigma$ uncertainty.}
\label{f.impact}
\end{figure}
\begin{figure}[t]
\includegraphics[width=0.48\textwidth]{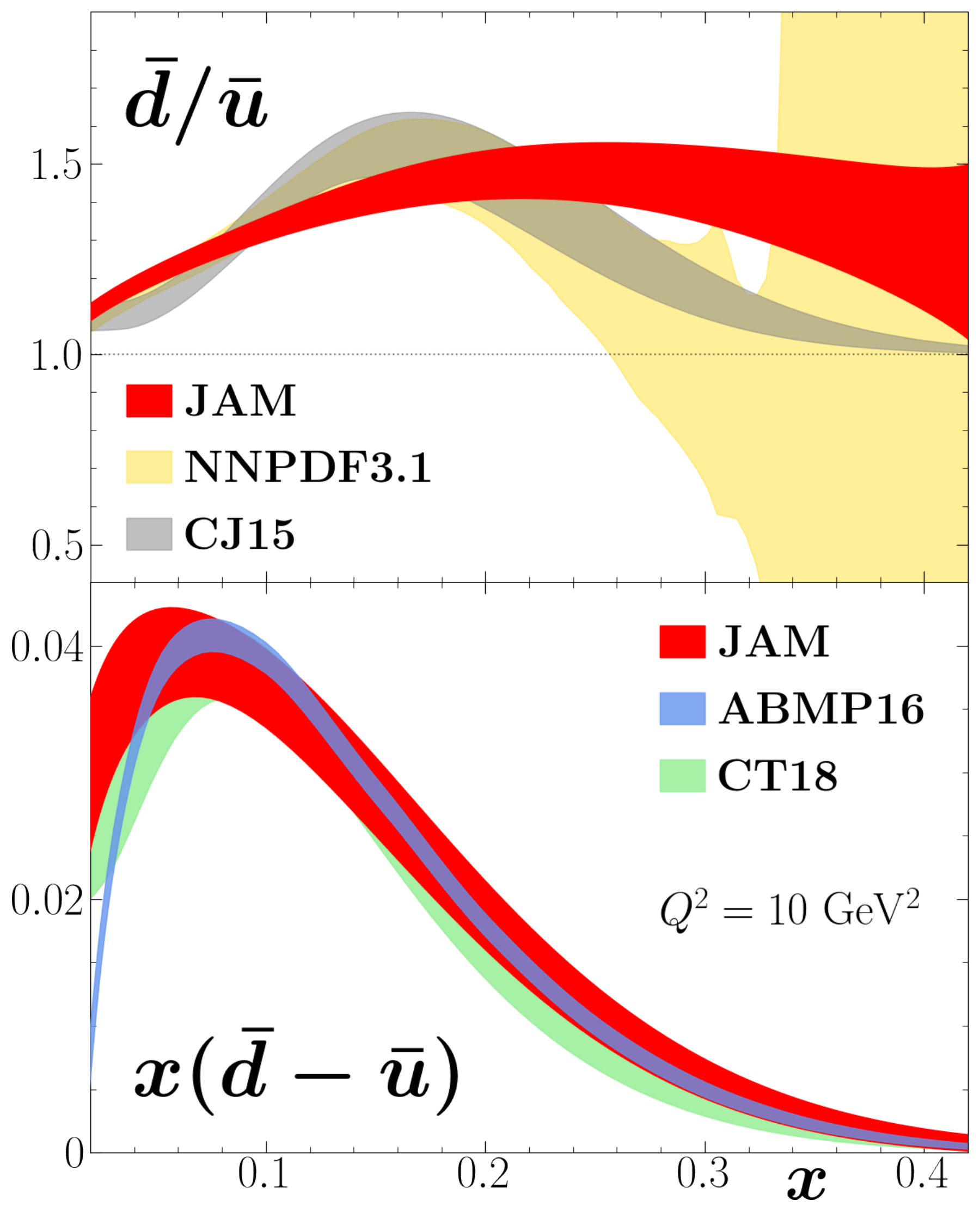}
\caption{Comparison of the JAM $\dbar/\ubar$ and $x(\dbar-\ubar)$ PDFs (red bands) with the NLO parametrizations from NNPDF3.1~\cite{NNPDF:2017mvq} (gold), ABMP16~\cite{Alekhin:2017kpj} (blue), CJ15~\cite{Accardi:2016qay} (gray), and CT18~\cite{Hou:2019efy} (green) at the scale $Q^2 = 10$~GeV$^2$.  All bands represent 1$\sigma$ uncertainty.}
\label{f.ratios}
\end{figure}

\section{Quality of Fit}
\label{sec.fit}

\begin{table}[b]
\caption{Summary of the $\chired$ values and number of data points for individual datasets used in this analysis.}
\begin{tabular}{l c r c}
\hhline{====}
process & ref. & $N_{\rm dat}$ & ~~$\chired$~       \\ 
\hline
DIS                         &         &         \\
~~~fixed target & \cite{BCDMS:1989qop, NewMuon:1996fwh, NewMuon:1996uwk, Whitlow:1991uw, JeffersonLabE00-115:2009jll, CLAS:2014jvt}
                            & ~~~2678~ & ~1.05    \\
~~~HERA & \cite{H1:2015ubc}
                            & 1185~    & ~1.27    \\ 
Drell-Yan                   &         &         \\
~~~NuSea~~~~~$pp$ & \cite{Webb:2003bj}
                            & 184~     & ~1.21    \\
~~~NuSea~~~~~$pD/2pp$ & \cite{NuSea:2001idv}
                            & 15~      & ~1.30    \\
~~~SeaQuest $pD/2pp$ & \cite{SeaQuest:2021zxb}
                            & 6~       & ~0.82    \\
$W$-lepton charge asymmetry                  &         &         \\
~~~STAR $W^+/W^-$ & \cite{STAR:2020vuq}
                            & 9~       & ~2.02    \\
~~~CMS & \cite{CMS:2011bet, CMS:2012ivw, CMS:2013pzl, CMS:2016qqr}
                            & 45~      & ~0.74    \\
~~~LHCb & \cite{LHCb:2014liz, LHCb:2016nhs}   
                            & 16~      & ~0.44    \\
$W$ rapidity & \cite{CDF:2009cjw, D0:2013lql}
                            & 27~      & ~1.18    \\
$Z$ rapidity & \cite{CDF:2010vek, D0:2007djv}
                            & 56~      & ~0.97    \\
inclusive jets & \cite{CDF:2007bvv, D0:2011jpq, STAR:2006opb}
                            & 200~     & ~1.11    \\
\hline
\textbf{total}  & & {\bf 4421}~ & {\bf ~1.12}\\
\hhline{====}
\end{tabular}
\label{t.chi2}
\end{table}

The datasets used in this analysis comprise inclusive DIS from hydrogen and deuterium, Drell-Yan lepton-pair production in $pp$ and $pD$ scattering, along with weak vector boson and jet production in $pp$ and $p\bar p$ collisions.
The various datasets are summarized in Table~\ref{t.chi2}, and their Born-level kinematics are displayed in \fref{f.kin}.
For DIS, we include $F_2$ structure function data from fixed target experiments from BCDMS~\cite{BCDMS:1989qop}, NMC~\cite{NewMuon:1996fwh, NewMuon:1996uwk}, SLAC~\cite{Whitlow:1991uw}, and Jefferson Lab~\cite{JeffersonLabE00-115:2009jll, CLAS:2014jvt}, as well as the reduced neutral and charged current proton cross sections from the combined H1/ZEUS analysis from HERA~\cite{H1:2015ubc}.
The cuts on the four-momentum transfer squared $Q^2$ and the hadronic final state masses $W$ are $Q^2 > m_c^2$ and $W^2 > 3$~GeV$^2$ for all DIS data.

Beyond DIS, we include Drell-Yan di-muon data from $pp$ and $pD$ collisions from the NuSea~\cite{Webb:2003bj, NuSea:2001idv} and SeaQuest~\cite{SeaQuest:2021zxb} experiments at Fermilab.
For weak vector boson mediated processes, we use $Z/\gamma^*$ cross sections and reconstructed $W$ asymmetries from the CDF and D0 Collaborations at the Tevatron~\cite{CDF:2010vek, D0:2007djv, CDF:2009cjw, D0:2013lql}; $W^\pm$-lepton asymmetries from CMS~\cite{CMS:2011bet, CMS:2012ivw, CMS:2013pzl, CMS:2016qqr} and LHCb~\cite{LHCb:2014liz, LHCb:2016nhs} at the LHC; and the ratio of $W^+$-lepton to $W^-$-lepton cross sections from STAR~\cite{STAR:2020vuq} at RHIC.
Finally, we include jet production data from CDF~\cite{CDF:2007bvv}, D0~\cite{D0:2011jpq}, and STAR~\cite{STAR:2006opb}.

The quality of the global fit is summarized in \tref{t.chi2}, which shows a global average $\chired=1.12$ for a total of $N_{\rm dat} = 4421$ data points.
The Drell-Yan data on the ratio $\sigDY_{pD}/2\sigDY_{pp}$ from both the NuSea and SeaQuest experiments are shown in \fref{f.SQ}.
Note that the data from the two experiments need not overlap, as they were taken at different kinematics with different values of $M_{\ell\ell}$, and for the ranges $0.36 < x_1 < 0.56$ for NuSea~\cite{NuSea:2001idv} and $0.48 < x_1 < 0.69$ for SeaQuest~\cite{SeaQuest:2021zxb}.
As suggested in Ref.~\cite{SeaQuest:2021zxb} and seen in \fref{f.SQ} (in combination with \eref{e.DYrat}), there is some tension between the new SeaQuest data and the earlier NuSea data, particularly above $x = 0.25$, where NuSea suggests $\dbar/\ubar < 1$ while SeaQuest indicates the opposite.
The amount of tension can be illustrated by removing the SeaQuest data, which then leads to an improvement in the $\chired$ for NuSea from 1.30 to 0.57.  
While the tension exists, we are nonetheless able to attain an acceptable description (within 2$\sigma$ theoretical uncertainty) of both datasets, as seen in \tref{t.chi2} and \fref{f.SQ}.

For the new $W$-lepton data from STAR, shown in \fref{f.STAR}, the description suffers slightly at high and at low pseudorapidities, leading to a $\chired$ of 2.02 for these data.
From Ref.~\cite{STAR:2020vuq}, it is known that this is a common feature of most PDF extractions.
This discrepancy is also partially due to some tension with the NuSea data, and the $\chired$ improves to 1.54 upon its removal.

\begin{figure*}[t]
\includegraphics[width=0.95\textwidth]{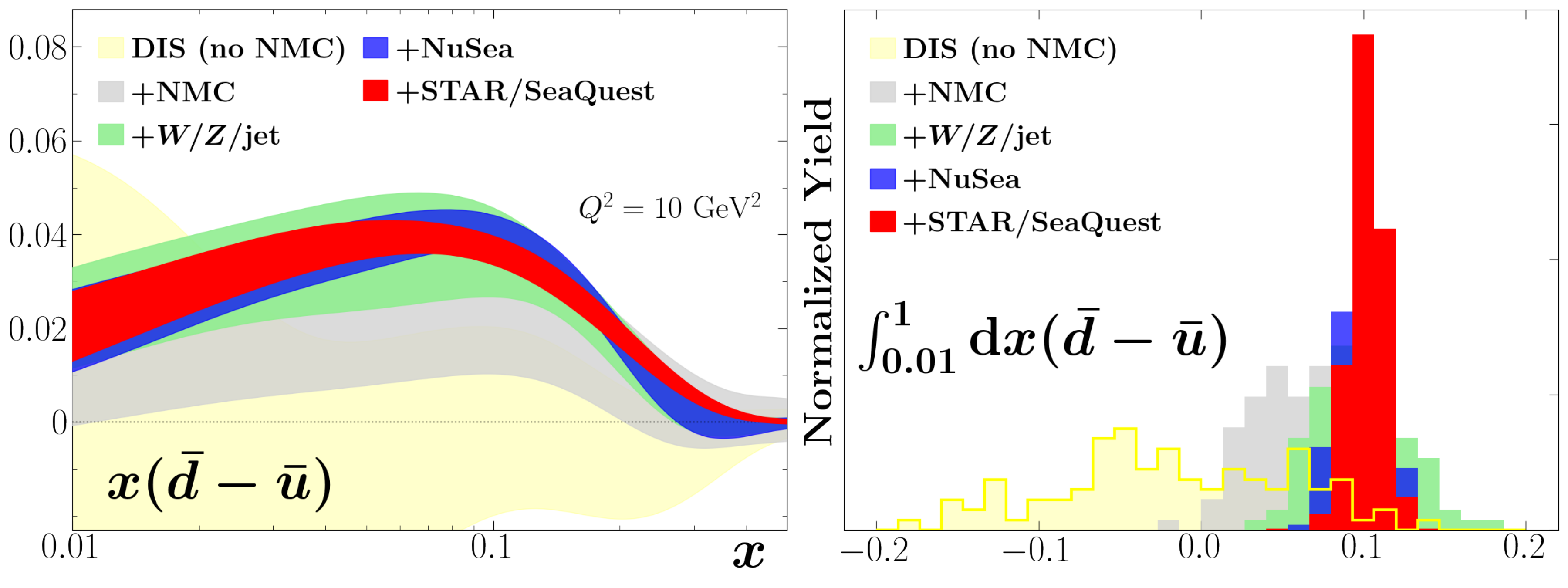}
\caption{
Comparison of $x(\dbar-\ubar)$ with different combinations of datasets (left panel).  First only DIS data~\cite{BCDMS:1989qop, Whitlow:1991uw, JeffersonLabE00-115:2009jll, CLAS:2014jvt, H1:2015ubc} excluding NMC (gold band).  Then data are added successively, with NMC~\cite{NewMuon:1996fwh, NewMuon:1996uwk} (gray); $W$-lepton and $W$ asymmetries/$Z$ boson production/jet production from RHIC~\cite{STAR:2006opb}, Tevatron~\cite{CDF:2010vek, D0:2007djv, CDF:2009cjw, D0:2013lql, CDF:2007bvv, D0:2011jpq}, and the LHC~\cite{CMS:2011bet, CMS:2012ivw, CMS:2013pzl, CMS:2016qqr, LHCb:2014liz, LHCb:2016nhs} (green); NuSea~\cite{Webb:2003bj, NuSea:2001idv} (blue); and finally the SeaQuest Drell-Yan~\cite{SeaQuest:2021zxb} and STAR $W$-lepton ratio~\cite{STAR:2020vuq} (red).
For the same combinations of data, the normalized yield of the Monte Carlo replicas for the truncated moment $\int_{0.01}^1 \diff x (\bar{d}-\bar{u})$ is also shown (right panel).}
\label{f.history}
\end{figure*}

\section{Extracted parton densities}
\label{sec.pdfs}

Our analysis is based on more than 900 Monte Carlo samples, which we use to ensure the statistical convergence of the PDFs, from which the means and expectation values are then computed using Eqs.~(\ref{e.EandVdef}).
The resulting parton densities are displayed in \fref{f.PDFs}, where we show the valence ($u_v$ and $d_v$), gluon ($g$), light antiquark ($\bar{d}+\bar{u}$), and strange ($s + \bar{s}$) distributions at a scale of $Q^2 = 10$~GeV$^2$.

In the valence sector, our results for both $u_v$ and $d_v$ are slightly larger than those from the other groups near the peaks of the distributions~\cite{NNPDF:2017mvq, Alekhin:2017kpj, Accardi:2016qay, Hou:2019efy}.
At lower $x$, the $u_v$ distribution agrees with the other groups, while the $d_v$ PDF agrees best with the CT18~\cite{Hou:2019efy} and NNPDF3.1~\cite{NNPDF:2017mvq} parametrizations.
For the gluon PDF, our results are largely in agreement with other extractions, although at low $x$ there are some differences with the ABMP16~\cite{Alekhin:2017kpj} fit.
The same is true for the sum of light sea quark distributions, $\bar{d}+\bar{u}$, but the disagreements are with CJ15~\cite{Accardi:2016qay}.
For the strange distribution $s+\bar{s}$, our result is somewhat suppressed at low $x$ relative to the other extractions.
This is consistent with previous JAM analyses~\cite{Sato:2019yez, Moffat:2021dji} that included semi-inclusive DIS cross section data, although these data are not included in the present analysis.
Note that this analysis does not include $Z$ boson production data from ATLAS, which were found \cite{ATLAS:2011qdp,ATLAS:2016nqi} to enhance the strange quark distribution at low $x$. Furthermore, we do not include a parametrization for the nonperturbative charm, but generate it entirely from radiation, which may also play a role in the shape of the strange quark PDF \cite{NNPDF:2017mvq}.

The impact on the light-quark sea asymmetry from both the new STAR measurement of $W$-lepton cross sections and the SeaQuest measurement of Drell-Yan di-muon production is shown in \fref{f.impact}.
From the baseline analysis, which excludes these new datasets, the STAR cross section ratios are added first in order to assess their impact.
While the STAR data do not lead to significant shifts in the central values of $\dbar/\ubar$, they do reduce the uncertainties somewhat, by up to $\approx 20\%$ at $x \approx 0.2$.
This modest impact can be understood from the fact that the NuSea Drell-Yan measurements are more directly sensitive to $\dbar/\ubar$ ({\it cf.} Eqs.~(\ref{e.DYrat}) and (\ref{e.Wrat})), and already provide the bulk of the constraints on the ratio even when compared to high precision $W$-lepton and reconstructed $W$ data from the Tevatron and LHC.
Since the STAR data overlap kinematically with the NuSea measurement, it is therefore difficult to improve on the extraction of $\dbar/\ubar$ using $W$-lepton production alone.

After adding the STAR data to the baseline, the SeaQuest Drell-Yan cross section ratios are then included.
In this case, the SeaQuest data greatly reduce the uncertainties on the $\dbar/\ubar$ ratio, by up to $\approx 50\%$ at high $x$, $x \gtrsim 0.3$.
As well as reducing the uncertainty, the addition of the SeaQuest data also increases the $\dbar/\ubar$ ratio for $x \approx 0.2$, where it remains above unity up to values of $x \approx 0.4$.
This is a direct consequence of the extended $x$ range of the  data compared with the earlier NuSea results, from $x \approx 0.3$ up to $x \approx 0.4$, with higher precision at the largest $x$ values. 
This feature is also reflected in the $\dbar-\ubar$ difference remaining positive across the entire range of $x$ probed.

With the new data from STAR and SeaQuest included, the final $\dbar/\ubar$ ratio and $x(\dbar-\ubar)$ difference are shown in \fref{f.ratios}, compared with the corresponding distributions from several other groups~\cite{NNPDF:2017mvq, Alekhin:2017kpj, Accardi:2016qay, Hou:2019efy} at $Q^2=10$~GeV$^2$ (see also Refs.~\cite{Park:2021kgf, Guzzi:2021fre}).
Our results are in agreement, within errors, with those from ABMP16~\cite{Alekhin:2017kpj} and CT18~\cite{Hou:2019efy} (except with ABMP16 at low $x \lesssim 0.04$), whose ratios also remain positive. 
Although there are differences in the shapes of the ratios, our results also largely agree within errors with those from NNPDF3.1~\cite{NNPDF:2017mvq} and CJ15~\cite{Accardi:2016qay}.
The disagreement with CJ15 at high $x$ results from their choice of parametrization that forces $\dbar/\ubar \to 1$ as $x \to 1$, and, more importantly, the fact that this fit predates the SeaQuest data, which pull the ratio upwards at large $x$.

It is instructive to examine the impact of individual datasets on the light-quark sea asymmetry, $x(\dbar-\ubar)$, which we illustrate in the left panel of \fref{f.history}.
Starting with inclusive DIS data only, and excluding data from the NMC experiment, the asymmetry is consistent with zero within large uncertainties.
Upon the inclusion of the NMC data~\cite{NewMuon:1996fwh, NewMuon:1996uwk}, the errors are significantly reduced, and the asymmetry gives an indication of deviation from zero in the range $0.01 < x < 0.2$.
When $W$-lepton, reconstructed $W$ and $Z$ boson, and jet production data from RHIC, Tevatron, and the LHC are included (but not the new STAR data~\cite{STAR:2020vuq}), the asymmetry becomes significantly larger, and more distinguishable from zero below $x=0.3$.
The new constraints come primarily from the high precision $W$ and $W$-lepton asymmetry measurements from the Tevatron and LHC, which are sensitive to $\bar{u}$ and $\bar{d}$ (see \eref{e.Wrat}).
The further addition of the NuSea~\cite{NuSea:2001idv} Drell-Yan data greatly decreases the uncertainty, showing that these data still provide a strong constraint on the asymmetry even when compared to the Tevatron and LHC $W$ and $W$-lepton asymmetries. 
Finally, the inclusion of the new SeaQuest~\cite{SeaQuest:2021zxb} and STAR~\cite{STAR:2020vuq} data reduces the uncertainty on the asymmetry even further, while increasing the magnitude at $x \gtrsim 0.2$,
as already seen in \fref{f.impact} except now displayed on a logarithmic scale.

The impact of the various datasets on the antiquark asymmetry can also be represented in the form of the truncated moment, $\int_{0.01}^1 \diff x (\bar{d}-\bar{u})$, illustrated in \fref{f.history} in the form of the normalized yield of the Monte Carlo replicas.
We choose $x=0.01$ for the lower limit as this is approximately the extent to which existing data provide information on the asymmetry (see \fref{f.kin}).
Note that because of large uncertainties associated with the small-$x$ extrapolation, estimates of the total moment are not as meaningful without  additional constraints on the $x \to 0$ behavior.
For the same combinations of datasets as described above, one observes that prior to the addition of the NMC data the truncated moments of the replicas can vary widely between $-0.2$ and $+0.15$.
Once the NMC data are added, the moments become almost entirely positive, and the yield continues to contract as more data are added. 
Once all of the data are included, the truncated moments are tightly gathered around 0.1.

\begin{figure}[t]
\includegraphics[width=0.48\textwidth]{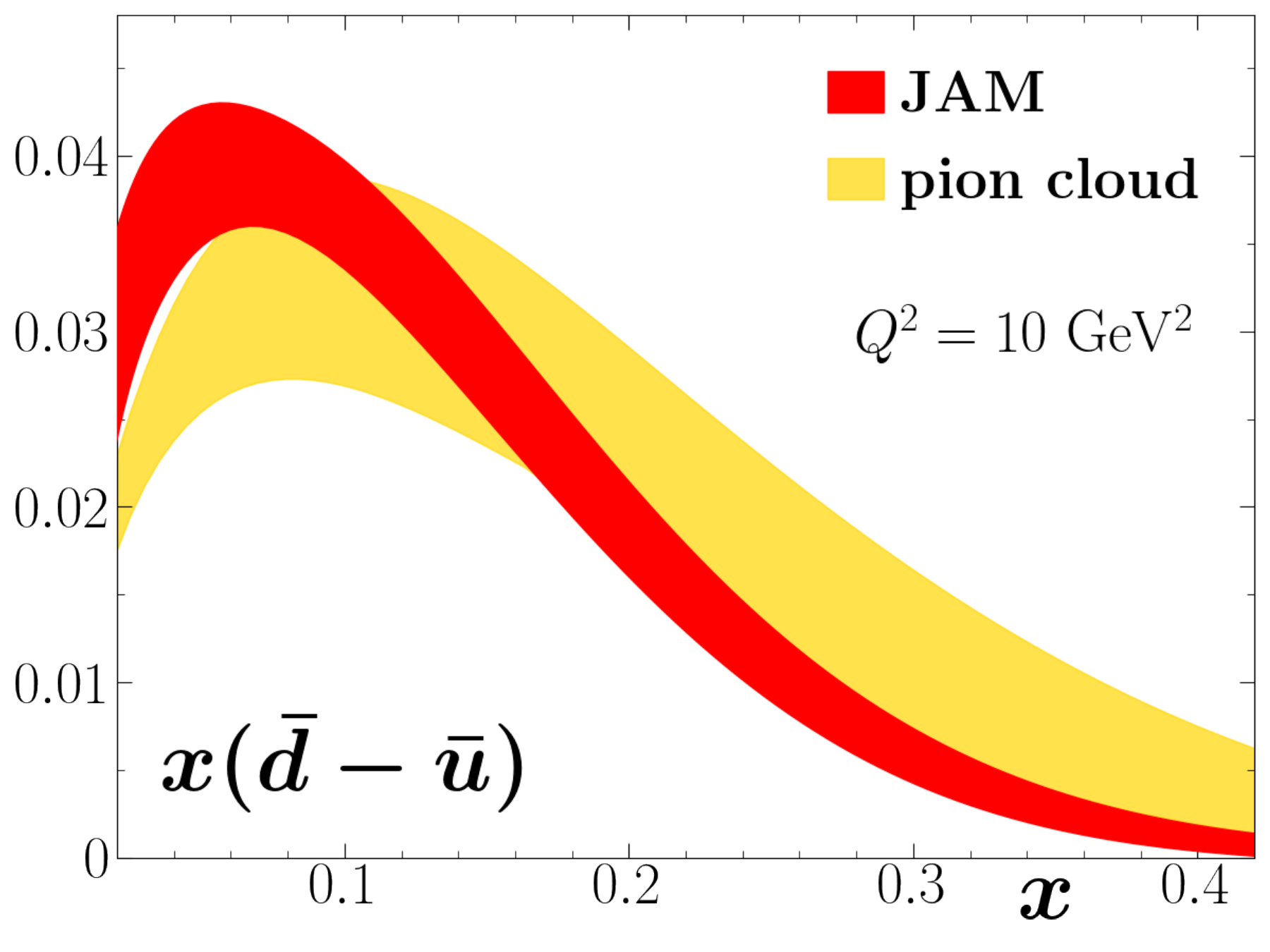}
\caption{Comparison of the extracted JAM $x(\dbar-\ubar)$ distribution (red band) at $Q^2 = 10$~GeV$^2$ with results from nonperturbative calculations based on chiral symmetry breaking and the pion cloud~\cite{Salamu:2014pka, Salamu:2019dok}. The JAM band represents 1$\sigma$ uncertainty, while the pion cloud band includes model dependence associated with the shape of the ultraviolet regulator function~\cite{McKenney:2015xis, Barry:2018ort, Barry:2021osv}.}
\label{f.models}
\end{figure}

The shape and magnitude of the $\dbar-\ubar$ asymmetry has long been an intriguing puzzle for our understanding of the nonperturbative structure of the nucleon.
The most common interpretation of the excess of $\bar d$ over $\bar u$ in the proton sea has been that associated with chiral symmetry breaking, and the consequent prevalence of the virtual $p \to n \pi^+$ dissociation~\cite{Thomas:1983fh}.
Scattering from the $\pi^+$ component of the proton wave function then naturally  enhances the $\bar d$ distribution, even though some of this will be cancelled by the subdominant $p \to \Delta^{++} \pi^-$ dissociation, which favors $\bar u$ over $\bar d$.
As an illustration, in \fref{f.models} we compare the inferred JAM asymmetry with $\bar d-\bar u$ at $x > 0$ calculated from a convolution of the $p \to$ baryon + $\pi$ splitting functions and the valence PDF of the pion~\cite{Salamu:2014pka, Salamu:2019dok, Burkardt:2012hk},
\begin{eqnarray}
(\bar d - \bar u)(x)
&=& \big[ \big(f_{n\pi^+} + f_{\Delta^0\pi^+} - f_{\Delta^{++}\pi^-}\big)
    \otimes \bar{q}_v^\pi
    \big](x), \qquad
\end{eqnarray}
where the convolution integral is defined as $f \otimes q = \int_0^1 \diff y \int_0^1 \diff z f(y)\, q(z)\, \delta(x-yz)$, and $y$ is the light-cone fraction of the proton's momentum carried by the pion.

For the calculation shown in Fig.~\ref{f.models}, the pion PDF is taken from the recent NLO analysis of pion-induced Drell-Yan and deep-inelastic leading neutron electroproduction data by Barry {\it et al.}~\cite{Barry:2021osv}.
The splitting functions are computed at one-pion loop order from chiral effective theory~\cite{Salamu:2014pka, Salamu:2019dok, Burkardt:2012hk}, using several different models for the ultraviolet regulators~\cite{McKenney:2015xis, Barry:2018ort}.
For the regulator mass parameters, we use the values from the global analysis in Ref.~\cite{Barry:2018ort}, for which the integrated splitting function was found to be $\langle n \rangle_{\pi N} = 0.22$.
Normalizing the various models of the regulator function, or hadronic form factor, to this value, the resulting band in Fig.~\ref{f.models} can be taken as a representation of the uncertainty on the calculated asymmetry.
The uncertainty also includes the errors from the Monte Carlo analysis, although these are small compared to the variations in the models.

The agreement between the calculated asymmetry and the extracted JAM PDFs is generally good within the uncertainties, with the model giving a slightly harder $x$ dependence compared with the global fit.
Qualitative agreement with the extracted asymmetry is also found for various other nonperturbative models (chiral loop and other) discussed in Refs.~\cite{Schreiber:1991qx, Henley:1990kw, Kumano:1991em, Melnitchouk:1991ui, Speth:1996pz, Kumano:1997cy, Melnitchouk:1998rv,  Pobylitsa:1998tk, Chang:2014jba, Geesaman:2018ixo, Alberg:2017ijg, Schreiber:1991tc, Bourrely:1994sc, Bourrely:1994nm, Steffens:1996bc}.

\section{Summary and outlook}
\label{sec.outlook}

Our global QCD analysis confirms the importance of the recent SeaQuest measurement of $pp$ and $pD$ Drell-Yan cross sections~\cite{SeaQuest:2021zxb}.
The inclusion of the SeaQuest data reduces the uncertainty on the $\dbar/\ubar$ ratio by up to 50\% at high values of $x$, and strongly suggests that the $\dbar-\ubar$ asymmetry remains positive up to $x \approx 0.4$. 
The impact from the latest $W^+/W^-$ STAR data~\cite{STAR:2020vuq} on the $\dbar/\ubar$ ratio, while still important, is less dramatic due to the lower sensitivity of $W$-lepton production data compared with the Drell-Yan data.

Although the new SeaQuest data indicate some tension with the earlier NuSea Drell-Yan measurement~\cite{NuSea:1998kqi, NuSea:2001idv} at large $x$, a good simultaneous description of both datasets is still possible due to the larger relative uncertainties of the NuSea data at high $x$.
The shape and magnitude of $\dbar-\ubar$ from the global analysis is consistent with expectations from nonperturbative models, such as those based on chiral symmetry breaking in QCD, all of which predict a positive asymmetry up to $x \sim 0.4-0.5$.
In the future, combining this analysis with semi-inclusive DIS data~\cite{Moffat:2021dji} could provide further constraints on the light-quark sea asymmetry in the proton.

\begin{acknowledgments}
We thank F.~Ringer and W.~Vogelsang for the code used for calculating the $W$-lepton cross sections.
We thank J.~J.~Ethier for his work on the $W$-lepton code and data collection, and P.~C.~Barry, M.~Diefenthaler, and Y.~Zhou for helpful discussions.
This work was supported by the U.S. Department of Energy Contract No.~DE-AC05-06OR23177, under which Jefferson Science Associates, LLC operates Jefferson Lab, and the National Science Foundation under grant number PHY-2110472.
The work of C.C. and A.M. was supported by the U.S. Department of Energy, Office of Science, Office of Nuclear Physics, within the framework of the TMD Topical Collaboration, and by Temple University (C.C.).
The work of N.S. was supported by the DOE, Office of Science, Office of Nuclear Physics in the Early Career Program.
\end{acknowledgments}



\begin{thebibliography}{99}

\bibitem{Gell-Mann:1964ewy}
M.~Gell-Mann,
\href{http://doi.org/10.1016/S0031-9163(64)92001-3}
{Phys. Lett. \textbf{8}, 214 (1964)}.

\bibitem{Zweig:1964ruk}
G.~Zweig,
\href{https://inspirehep.net/literature/11881}
{preprint CERN-TH-401}.

\bibitem{Feynman:1973xc}
R.~P.~Feynman,
\href{https://inspirehep.net/literature/85512}
{\it Photon-hadron interactions}
(Reading, Massachusetts, 1972).

\bibitem{Geesaman:2018ixo}
D.~F.~Geesaman and P.~E.~Reimer,
\href{https://doi.org/10.1088/1361-6633/ab05a7}
{Rep. Prog. Phys. \textbf{82}, 046301 (2019)}.

\bibitem{Ross:1978xk}
D.~A.~Ross and C.~T.~Sachrajda,
\href{http://doi.org/10.1016/0550-3213(79)90004-X}
{Nucl. Phys. \textbf{B149}, 497 (1979)}.

\bibitem{Thomas:1983fh}
A. W. Thomas,
\href{http://doi.org/10.1016/0370-2693(83)90026-6}
{Phys. Lett. B \textbf{126}, 97 (1983)}.

\bibitem{Schreiber:1991qx}
A.~W.~Schreiber, P.~J.~Mulders, A.~I.~Signal, and A.~W.~Thomas,
\href{http://doi.org/10.1103/PhysRevD.45.3069}
{Phys. Rev. D \textbf{45}, 3069 (1992)}.

\bibitem{Henley:1990kw}
E.~M.~Henley and G.~A.~Miller,
\href{http://doi.org/10.1016/0370-2693(90)90735-O}
{Phys. Lett. B \textbf{251}, 453 (1990)}.

\bibitem{Kumano:1991em}
S.~Kumano and J.~T.~Londergan,
\href{http://doi.org/10.1103/PhysRevD.44.717}
{Phys. Rev. D \textbf{44}, 717 (1991)}.

\bibitem{Melnitchouk:1991ui}
W.~Melnitchouk, A.~W.~Thomas and A.~I.~Signal,
\href{http://doi.org/10.1007/BF01284484}
{Z.~Phys. A \textbf{340}, 85 (1991)}.

\bibitem{Speth:1996pz}
J.~Speth and A.~W.~Thomas,
\href{http://doi.org/10.1007/0-306-47073-X\_2}
{Adv. Nucl. Phys. \textbf{24}, 83 (1997)}.

\bibitem{Kumano:1997cy}
S.~Kumano,
\href{http://doi.org/10.1016/S0370-1573(98)00016-7}
{Phys. Rep. \textbf{303}, 183 (1998)}.

\bibitem{Melnitchouk:1998rv}
W.~Melnitchouk, J.~Speth and A.~W.~Thomas,
\href{http://doi.org/10.1103/PhysRevD.59.014033}
{Phys. Rev. D \textbf{59}, 014033 (1998)}.

\bibitem{Pobylitsa:1998tk}
P.~V.~Pobylitsa, M.~V.~Polyakov, K.~Goeke, T.~Watabe and C.~Weiss,
\href{https://doi.org/10.1103/PhysRevD.59.034024}
{Phys. Rev. D \textbf{59}, 034024 (1999)}.

\bibitem{Chang:2014jba}
W.~C.~Chang and J.~C.~Peng,
\href{https://doi.org/10.1016/j.ppnp.2014.08.002}
{Prog. Part. Nucl. Phys.~\textbf{79}, 95 (2014)}.

\bibitem{Alberg:2017ijg}
M. Alberg and G. A. Miller, 
\href{http://doi.org/10.1103/PhysRevC.100.035205}
{Phys. Rev. C \textbf{100}, 035205 (2019)}.

\bibitem{Field:1976ve}
R.~D.~Field and R.~P.~Feynman,
\href{http://doi.org/10.1103/PhysRevD.15.2590}
{Phys. Rev. D \textbf{15}, 2590 (1977)}.

\bibitem{Schreiber:1991tc}
A.~W.~Schreiber, A.~I.~Signal and A.~W.~Thomas,
\href{http://doi.org/10.1103/PhysRevD.44.2653}
{Phys. Rev. D \textbf{44}, 2653 (1991)}.

\bibitem{Bourrely:1994sc}
C.~Bourrely and J.~Soffer,
\href{https://doi.org/10.1016/0550-3213(94)90137-6}
{Nucl. Phys. \textbf{B423}, 329 (1994)}.

\bibitem{Bourrely:1994nm}
C.~Bourrely and J.~Soffer,
\href{http://doi.org/10.1103/PhysRevD.51.2108}
{Phys. Rev. D \textbf{51}, 2108 (1995)}.

\bibitem{Steffens:1996bc}
F.~M.~Steffens and A.~W.~Thomas,
\href{http://doi.org/10.1103/PhysRevC.55.900}
{Phys. Rev. C \textbf{55}, 900 (1997)}.

\bibitem{Broadhurst:2004jx}
D.~J.~Broadhurst, A.~L.~Kataev and C.~J.~Maxwell,
\href{https://doi.org/10.1016/j.physletb.2004.03.059}
{Phys. Lett. B \textbf{590} (2004), 76}.

\bibitem{Ji:2013dva}
X.~Ji,
\href{https://doi.org/10.1103/PhysRevLett.110.262002}
{Phys. Rev. Lett. \textbf{110}, 262002 (2013)}.

\bibitem{Ma:2014jla}
Y.~Q.~Ma and J.-W.~Qiu,
\href{https://doi.org/10.1103/PhysRevD.98.074021}
{Phys. Rev. D \textbf{98}, 074021 (2018)}.

\bibitem{Orginos:2017kos}
K.~Orginos, A.~Radyushkin, J.~Karpie and S.~Zafeiropoulos,
\href{https://doi.org/10.1103/PhysRevD.96.094503}
{Phys. Rev. D \textbf{96}, 094503 (2017)}.

\bibitem{Chen:2017mzz}
J.~W.~Chen, T.~Ishikawa, L.~Jin, H.~W.~Lin, Y.~B.~Yang, J.~H.~Zhang and Y.~Zhao,
\href{https://doi.org/10.1103/PhysRevD.97.014505}
{Phys. Rev. D \textbf{97}, 014505 (2018)}.

\bibitem{Alexandrou:2021oih}
C.~Alexandrou, M.~Constantinou, K.~Hadjiyiannakou, K.~Jansen and F.~Manigrasso,
\href{https://arxiv.org/abs/2106.16065}
{arXiv:2106.16065 [hep-lat]}.

\bibitem{Ito:1980ev}
A.~S.~Ito \textit{et al.}, 
\href{https://doi.org/10.1103/PhysRevD.23.604}
{Phys. Rev. D \textbf{23}, 604 (1981)}.

\bibitem{NewMuon:1991hlj}
P. Amaudruz \textit{et al.},
\href{http://doi.org/10.1103/PhysRevLett.66.2712}
{Phys. Rev. Lett. \textbf{66}, 2712 (1991)}.

\bibitem{NewMuon:1993oys}
M. Arneodo \textit{et al.},
\href{http://doi.org/10.1103/PhysRevD.50.R1}
{Phys. Rev. D \textbf{50}, R1 (1994)}.

\bibitem{Gottfried:1967kk}
K. Gottfried,
\href{https://doi.org/10.1103/PhysRevLett.18.1174}
{Phys. Rev. Lett. {\bf 18}, 1174 (1967)}.

\bibitem{NA51:1994xrz}
A.~Baldit \textit{et al.},
\href{https://doi.org/10.1016/0370-2693(94)90884-2}
{Phys.~Lett.~B~\textbf{332}, 244 (1994)}.

\bibitem{HERMES:1998uvc}
K.~Ackerstaff \textit{et al.},
\href{https://doi.org/10.1103/PhysRevLett.81.5519}
{Phys. Rev. Lett. \textbf{81}, 5519 (1998)}.

\bibitem{Webb:2003bj}
J. Webb, Ph.D. Thesis, New Mexico State University (2002),
\href{https://arxiv.org/abs/hep-ex/0301031}
{arXiv:hep-ex/0301031}.

\bibitem{NuSea:1998kqi}
E.~A.~Hawker \textit{et al.},
\href{https://doi.org/10.1103/PhysRevLett.80.3715}
{Phys. Rev. Lett. {\bf 80}, 3715 (1998)}.

\bibitem{NuSea:2001idv}
R.~S.~Towell \textit{et al.},
\href{https://doi.org/10.1103/PhysRevD.64.052002}
{Phys. Rev. D \textbf{64}, 052002 (2001)}.

\bibitem{Ellis:1990ti}
S.~D.~Ellis and W.~J.~Stirling,
\href{https://doi.org/10.1016/0370-2693(91)90684-I}
{Phys. Lett. B \textbf{256}, 258 (1991)}.

\bibitem{SeaQuest:2021zxb}
J.~Dove \textit{et al.},
\href{https://doi.org/10.1038/s41586-021-03282-z}
{Nature \textbf{590}, 561 (2021)}.

\bibitem{STAR:2020vuq}
J.~Adam \textit{et al.},
\href{https://doi.org/10.1103/PhysRevD.103.012001}
{Phys. Rev. D \textbf{103}, 012001 (2021)}.

\bibitem{Becher:2007ty}
T. Becher, M. Neubert and G. Xu,
\href{ https://doi.org/10.1088/1126-6708/2008/07/030}
{JHEP {\bf 07}, 030 (2008)}.

\bibitem{Ehlers:2014jpa}
P.~J.~Ehlers, A.~Accardi, L.~T.~Brady and W.~Melnitchouk,
\href{https://doi.org/10.1103/PhysRevD.90.014010}
{Phys. Rev. D \textbf{90}, 014010 (2014)}.

\bibitem{Ringer:2015oaa}
F. Ringer and W. Vogelsang,
\href{https://doi.org/10.1103/PhysRevD.91.094033}
{Phys. Rev. D {\bf 91}, 094033 (2015)}.

\bibitem{Ebert:2020dfc}
M.~Ebert, J.~Michel, I.~Stewart and F.~J.~Tackmann,
\href{https://doi.org/10.1007/JHEP04(2021)102}
{JHEP \textbf{04}, 102 (2021)}.

\bibitem{Moffat:2019qll}
E.~Moffat, T.~C.~Rogers, W.~Melnitchouk, N.~Sato and F.~Steffens,
\href{https://doi.org/10.1103/PhysRevD.99.096008}
{Phys. Rev. D \textbf{99}, 096008 (2019)}.

\bibitem{Cocuzza:2021DIS}
C. Cocuzza, W. Melnitchouk, A. Metz and N. Sato,
in preparation (2021). 

\bibitem{Gribov:1972ri}
V.~N.~Gribov and L.~N.~Lipatov,
\href{https://inspirehep.net/literature/73449}
{Sov. J. Nucl. Phys. \textbf{15}, 438 (1972)}.

\bibitem{Dokshitzer:1977sg}
Y.~L.~Dokshitzer,
\href{https://inspirehep.net/literature/126153}
{Sov. Phys. JETP \textbf{46}, 641 (1977)}.

\bibitem{Altarelli:1977zs}
G.~Altarelli and G.~Parisi,
\href{https://doi.org/10.1016/0550-3213(77)90384-4}
{Nucl. Phys. \textbf{B126}, 298 (1977)}.

\bibitem{ParticleDataGroup:2018ovx}
M. Tanabashi \textit{et al.} (Particle Data Group),
\href{https://doi.org/10.1103/PhysRevD.98.030001}
{Phys. Rev. D {\bf 98}, 030001 (2018), and 2019 update}.

\bibitem{Sato:2016tuz}
N. Sato, W. Melnitchouk, S. E. Kuhn, J. J. Ethier and A. Accardi,
\href{http://doi.org/10.1103/PhysRevD.93.074005}
{Phys. Rev. D \textbf{93}, 074005 (2016)}.

\bibitem{Sato:2016wqj}
N.~Sato, J.~J.~Ethier, W.~Melnitchouk, M.~Hirai, S.~Kumano and A.~Accardi,
\href{http://doi.org/10.1103/PhysRevD.94.114004}
{Phys. Rev. D \textbf{94}, 114004 (2016)}.

\bibitem{Ethier:2017zbq}
J.~J.~Ethier, N.~Sato and W.~Melnitchouk,
\href{http://doi.org/10.1103/PhysRevLett.119.132001}
{Phys. Rev. Lett. \textbf{119}, 132001 (2017)}.

\bibitem{Sato:2019yez}
N. Sato, C. Andres, J. J. Ethier and W. Melnitchouk,
\href{http://doi.org/10.1103/PhysRevD.101.074020}
{Phys. Rev. D \textbf{101}, 074020 (2020)}.

\bibitem{Moffat:2021dji}
E.~Moffat, W.~Melnitchouk, T.~Rogers and N.~Sato,
\href{https://doi.org/10.1103/PhysRevD.104.016015}
{Phys. Rev. D \textbf{104}, 016015 (2021)}.

\bibitem{BCDMS:1989qop}
A. C. Benvenuti \textit{et al.},
\href{https://doi.org/10.1016/0370-2693(89)91637-7}
{Phys. Lett. B {\bf 223}, 485 (1989)}; 
{\it ibid.} 
\href{https://doi.org/10.1016/0370-2693(90)91231-Y}
{B {\bf 237}, 592 (1990)}.

\bibitem{NewMuon:1996fwh}
M. Arneodo \textit{et al.},
\href{https://doi.org/10.1016/S0550-3213(96)00538-X}
{Nucl. Phys. {\bf B483}, 3 (1997)}.

\bibitem{NewMuon:1996uwk}
M. Arneodo \textit{et al.},
\href{https://doi.org/10.1016/S0550-3213(96)00673-6}
{Nucl. Phys. {\bf B487}, 3 (1997)}.

\bibitem{Whitlow:1991uw}
L. W. Whitlow \textit{et al.},
\href{https://doi.org/10.1016/0370-2693(92)90672-Q}
{Phys. Lett. B {\bf 282}, 475 (1992)}.

\bibitem{JeffersonLabE00-115:2009jll}
S. P. Malace \textit{et al.},
\href{https://doi.org/10.1103/PhysRevC.80.035207}
{Phys. Rev. C {\bf 80}, 035207 (2009)}.

\bibitem{CLAS:2014jvt}
S.~Tkachenko \textit{et al.},
\href{https://doi.org/10.1103/PhysRevC.89.045206}
{Phys. Rev. C \textbf{89}, 045206 (2014)}.

\bibitem{H1:2015ubc}
H. Abramowicz \textit{et al.},
\href{https://doi.org/10.1140/epjc/s10052-015-3710-4}
{Eur. Phys. J. C {\bf 75}, 580 (2015)}.

\bibitem{CDF:2010vek}
T. Aaltonen \textit{et al.},
\href{https://doi.org/10.1016/j.physletb.2010.06.043}
{Phys. Lett. B {\bf 692}, 232 (2010)}.

\bibitem{D0:2007djv}
V. M. Abazov \textit{et al.},
\href{https://doi.org/10.1103/PhysRevD.76.012003}
{Phys. Rev. D {\bf 76}, 012003 (2007)}.

\bibitem{CDF:2009cjw}
T. Aaltonen {\it et al.},
\href{https://doi.org/10.1103/PhysRevLett.102.181801}
{Phys. Rev. Lett. {\bf 102}, 181801 (2009)}.

\bibitem{D0:2013lql}
V. M. Abazov {\it et al.},
\href{https://doi.org/10.1103/PhysRevLett.112.151803}
{Phys. Rev. Lett. {\bf 112}, 151803 (2014)}; 
{\it ibid.} 
\href{https://doi.org/10.1103/PhysRevLett.114.049901}
{{\bf 114}, 049901 (2015)}.

\bibitem{CMS:2011bet}
S. Chatrchyan \textit{et al.},
\href{ https://doi.org/10.1007/JHEP04(2011)050}
{JHEP {\bf 04}, 050 (2011)}.

\bibitem{CMS:2012ivw}
S. Chatrchyan \textit{et al.},
\href{https://doi.org/10.1103/PhysRevLett.109.111806}
{Phys. Rev. Lett. {\bf 109}, 111806 (2012)}.

\bibitem{CMS:2013pzl}
S. Chatrchyan \textit{et al.},
\href{https://doi.org/10.1103/PhysRevD.90.032004}
{Phys. Rev. D {\bf 90}, 032004 (2014)}.

\bibitem{CMS:2016qqr}
V. Khachatryan \textit{et al.},
\href{https://doi.org/10.1140/epjc/s10052-016-4293-4}
{Eur. Phys. J. C {\bf 76}, 469 (2016)}.

\bibitem{LHCb:2014liz}
R.~Aaij \textit{et al.},
\href{https://doi.org/10.1007/JHEP12(2014)079}
{JHEP \textbf{12}, 079 (2014)}.

\bibitem{LHCb:2016nhs}
R.~Aaij \textit{et al.},
\href{https://doi.org/10.1007/JHEP01(2016)155}
{JHEP \textbf{01}, 155 (2016)}.

\bibitem{CDF:2007bvv}
A. Abulencia \textit{et al.},
\href{https://doi.org/10.1103/PhysRevD.75.092006}
{Phys. Rev. D {\bf 75}, 092006 (2007)}
[Erratum: 
\href{https://doi.org/10.1103/PhysRevD.75.119901}
{Phys. Rev. D {\bf 75}, 119901 (2007)}].

\bibitem{D0:2011jpq}
V. M. Abazov \textit{et al.},
\href{https://doi.org/10.1103/PhysRevLett.101.062001}
{Phys. Rev. Lett. {\bf 101}, 062001 (2008)}.

\bibitem{STAR:2006opb}
B. I. Abelev \textit{et al.},
\href{https://doi.org/10.1103/PhysRevLett.97.252001}
{Phys. Rev. Lett. {\bf 97}, 252001 (2006)}.

\bibitem{NNPDF:2017mvq}
R. D. Ball \textit{et al.},
\href{https://doi.org/10.1140/epjc/s10052-017-5199-5}
{Eur. Phys. J. C {\bf 77}, 663 (2017)}.

\bibitem{Alekhin:2017kpj}
S. Alekhin, J. Bl\"{u}mlein, S. Moch and R. Pla\v{c}akyt\.{e},
\href{https://doi.org/10.1103/PhysRevD.96.014011}
{Phys. Rev. D {\bf 96}, 014011 (2017)}.

\bibitem{Accardi:2016qay}
A. Accardi, L.T. Brady, W. Melnitchouk, J. F. Owens and N. Sato,
\href{https://doi.org/10.1103/PhysRevD.93.114017}
{Phys. Rev. D {\bf 93}, 114017 (2016)}.

\bibitem{Hou:2019efy}
T. Hou \textit{et al.},
\href{https://doi.org/10.1103/PhysRevD.103.014013}
{Phys. Rev. D {\bf 103}, 014013 (2021)}.

\bibitem{ATLAS:2011qdp}
G.~Aad \textit{et al.}, 
\href{https://doi.org/10.1103/PhysRevD.85.072004}
{Phys. Rev. D \textbf{85}, 072004 (2012)}.

\bibitem{ATLAS:2016nqi}
M.~Aaboud \textit{et al.}, 
\href{https://doi.org/10.1140/epjc/s10052-017-4911-9}
{Eur. Phys. J. C \textbf{77}, 367 (2017)}.

\bibitem{Park:2021kgf}
S.~Park, A.~Accardi, X.~Jing and J.~F.~Owens,
\href{https://arxiv.org/abs/2108.05786}
{arXiv:hep-ex/2108.057866}.

\bibitem{Guzzi:2021fre}
M.~Guzzi \textit{et al.},
\href{https://arxiv.org/abs/2108.06596}
{arXiv:hep-ex/2108.06596}.

\bibitem{Barry:2021osv}
P.~C.~Barry, C.-R.~Ji, N.~Sato and W.~Melnitchouk,
\href{https://arxiv.org/abs/2108.05822}
{arXiv:2108.05822 [hep-ph]}.

\bibitem{Salamu:2014pka}
Y.~Salamu, C.-R.~Ji, W.~Melnitchouk and P.~Wang,
\href{https://doi.org/10.1103/PhysRevLett.114.122001}
{Phys. Rev. Lett. \textbf{114}, 122001 (2015)}.

\bibitem{Salamu:2019dok}
Y.~Salamu, C.-R.~Ji, W.~Melnitchouk, A.~W.~Thomas, P.~Wang and X.~G.~Wang,
\href{https://doi.org/10.1103/PhysRevD.100.094026}
{Phys. Rev. D \textbf{100}, 094026 (2019)}.

\bibitem{Burkardt:2012hk}
M.~Burkardt, K.~S.~Hendricks, C.-R.~Ji, W.~Melnitchouk and A.~W.~Thomas,
\href{https://doi.org/10.1103/PhysRevD.87.056009}
{Phys. Rev. D \textbf{87}, 056009 (2013)}.

\bibitem{Barry:2018ort}
P.~C.~Barry, N.~Sato, W.~Melnitchouk and C.-R.~Ji,
\href{https://doi.org/10.1103/PhysRevLett.121.152001}
{Phys. Rev. Lett. \textbf{121}, 152001 (2018)}.

\bibitem{McKenney:2015xis}
J.~R.~McKenney, N.~Sato, W.~Melnitchouk and C.-R.~Ji,
\href{https://doi.org/10.1103/PhysRevD.93.054011}
{Phys. Rev. D \textbf{93}, 054011 (2016)}.


\end{thebibliography}
\end{document}